\documentclass
[prb,aps,twocolumn,showpacs,superscriptaddress,floatfix]{revtex4}%
\usepackage{epsfig}
\usepackage{amsmath}
\usepackage{amssymb}
\usepackage{amsfonts}
\usepackage{mathptmx}
\usepackage{eucal}
\usepackage{bm}%
\usepackage{graphicx}
\providecommand{\U}[1]{\protect\rule{.1in}{.1in}}
\providecommand{\U}[1]{\protect\rule{.1in}{.1in}}
\begin{document}
\title{Spatially-resolved probing of a non-equilibrium superconductor}
\author{K. Yu. Arutyunov}
\email{Konstantin.Arutyunov@phys.jyu.fi}
\affiliation{University of Jyv\"askyl\"a, Department of Physics, NanoScience Centre, PB 35,
40014 Jyv\"askyl\"a, Finland}
\affiliation{Nuclear Physics Institute, Moscow State University, 119992 Moscow, Russia}
\author{H.-P. Auraneva}
\affiliation{University of Jyv\"askyl\"a, Department of Physics, NanoScience Centre, PB 35,
40014 Jyv\"askyl\"a, Finland}
\author{A.~S.~Vasenko}
\affiliation{Institut Laue-Langevin, 6 rue Jules Horowitz, BP 156, 38042 Grenoble, France}
\affiliation{LPMMC, Universit\'{e} Joseph Fourier and CNRS, 25 Avenue des Martyrs, BP 166,
38042 Grenoble, France}
\affiliation{Donostia International Physics Center (DIPC), Manuel de Lardizbal 5, E-20018
San Sebasti\'{a}n, Spain}
\date{\today}

\begin{abstract}
Spatially resolved relaxation of non-equilibrium quasiparticles in a
superconductor at ultra-low temperatures was experimentally studied. It was
found that the quasiparticle injection through a tunnel junction results in
the modification of the shape of the I-V characteristic of a remote `detector'
junction. The effect depends on the temperature, injection current and
proximity to the injector. The phenomena can be understood in terms of
creation of quasiparticle charge and energy disequilibrium characterized by
two different length scales $\Lambda_{Q^{\ast}}$ $\sim5$ $\mu$m and
$\Lambda_{T^{\ast}}\sim$ $40$ $\mu$m. The findings are in good agreement with
existing phenomenological models, while more elaborated microscopic theory is
mandatory for detailed quantitative comparison with experiment. The results
are of fundamental importance for understanding electron transport phenomena
in various nanoelectronic circuits.

\end{abstract}

\pacs{74.25.Fy, 74.40.+k, 74.50.+r, 74.78.Na}
\maketitle



\section{Introduction}

Conversion of electric current at an interface between different materials is
a common process in any electric circuit. At nanoscales the whole system might
behave as an interface if the dimension(s) are comparable to the
characteristic relaxation length. Of particular interest are boundaries with a
superconductor where electric current converts from single electrons to
Cooper pairs. In a superconductor at a finite temperature there are always
non-paired electrons called \textit{equilibrium }quasiparticles. In the
presence of additional disturbance their concentration can be increased by
\textit{non-equilibrium} quasiparticles. If these excitations originate solely
from pair-breaking they equally populate excitation spectrum with momenta
$p_{Q}>p_{F}$ (electron-like) and $p_{Q}<p_{F}$ (hole-like), where $p_{F}$ is
the Fermi momentum. Such a perturbation contributes to the
\textit{longitudinal} (or energy) mode (L) contrary to \textit{transverse} (or
charge imbalance) mode (T) where the two branches of the excitation spectrum
are non-equally populated. \cite{Schmid-Schoen} Within the Keldysh formalism
one can define the non-equilibrium distribution functions $f_{L,T}$.
\cite{Belzig} The usual electron distribution function is $2f=1-f_{L}-f_{T}$.

These phenomena have attracted attention in mid-70s resulting in an impressive
number of papers. \cite{Gray book, Kopnin book} Those early experiments were
mainly performed on sandwich-type flat structures not adequate for
spatially-resolved studies. Agreement between experiment and theory was
established reliably mainly in the high temperature limit $T\rightarrow T_{c}%
$. The opposite limit $T \ll T_{c}$ has been poorly investigated. Recently the
interest to the problem increased, \cite{Yagi, Pekola energy relaxation,
Kopnin arxive, Beckmann Lambda*} while the understanding is still far from
being satisfactory. In addition to the general interest in physics of
superconductivity, relaxation of quasiparticle excitations is a bottleneck
process in many nanoelectronic applications: solid state coolers, \cite{Pekola
NIS review, Arutyunov_cooling, Vasenko_cooling} 
cold/hot electron bolometers, \cite{Kuzmin bolometer} and Cooper
pair boxes. \cite{SET poisoning} Experimental study of the corresponding
phenomena is the subject of the paper.

The paper is organized as follows. In the next Section, we describe our setup
and the experimental conditions. In Sec.~\ref{results} we discuss our
experimental data. We also formulate a phenomenological model which is in a
good agreement with our findings. The model enables us to obtain the energy
and charge modes relaxation lengths. Finally, we summarize the results in
Sec.~\ref{secconcl}.

\begin{figure}[t]
\epsfxsize=7cm\epsffile{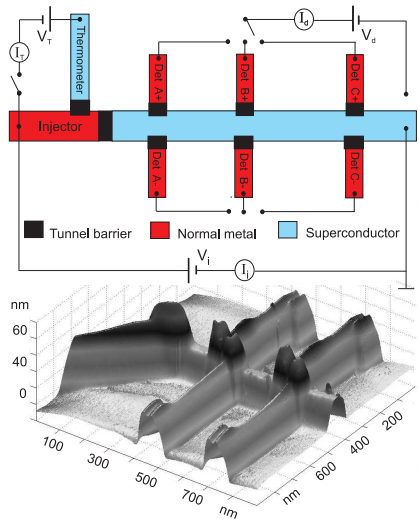}\caption{(Color online) Top: schematics of
the sample and measurements. Electron temperature of the normal metal injector
is measured with the NIS junction `Thermometer'. The quasiparticle relaxation
in the superconducting long bar is determined by the `Detectors': either a
single NIS junction, or a pair of connected in series NISIN junctions from the
opposite sides of the superconductor. Bottom: scanning probe microscope image
of the Si/SiO$_{x}$ substrate with two closest detector pairs and the injector
(left most massive probe). Typical thickness of the aluminum wire, copper
detectors, and injector are 25 , 40 and 80 nm, and the width 400, 180 and 1600
nm, respectively. Critical temperature of the aluminum wire is about 1.35 K
and the low temperature mean free path $\ell\simeq$20 nm.}%
\label{Fig-1}%
\end{figure}

\section{Experiment}

\label{experiment}

Multi-terminal nanostructures (Fig.~\ref{Fig-1}) were fabricated using
electron beam lithography and UHV evaporation of aluminum (S) and copper (N)
separated by naturally-grown aluminum oxide (I). The sample layout is
conceptually close to the layout, described in Refs.~\onlinecite{Yagi} and
\onlinecite{Yu-Mercereau-Dolan-Jackel}. Electrons were injected into the
superconductor through a large-area NIS injector with typical tunnel
resistance $\sim$ 3 k$\Omega$. The induced disequilibrium was measured by
remote detector junctions with typical tunnel resistance $\sim$ 50 k$\Omega$.
Even at the lowest values of the utilized tunnel resistances one can neglect
the proximity effect.

The $^{3}$He$^{4}$He dilution refrigerator was located inside the
electromagnetically shielded room with only battery-powered analogue
pre-amplifiers inside. Typical measuring set-up (Fig.~\ref{Fig-1}, top panel)
consisted of two circuits: injector and detector. For detection two
complementary configurations were used: either a single NIS junction, or two
connected in series NISIN junctions from both sides of the superconductor. The
latter was found to be less sensitive to dc offsets presumably originating
from parasitic thermoelectric effects. Care was taken not to increase the
temperature of the system by the Joule heating at the injector by making the
electrode rather massive (Fig.~\ref{Fig-1}, bottom panel). An extra NIS junction
(marked `Thermometer' in Fig.~\ref{Fig-1} and located $\simeq$1 $\mu$m away
from the injector junction) was used to measure the electron temperature of
the injector $T_{e}^{i}(I_{i})$ by fitting the experimental $I_{T}(V_{T}%
,I_{i}=\operatorname{const})$ dependencies with the text-book expression for
the tunnel current of a NIS junction. Here and below $I_{i}(V_{i})$,
$I_{d}(V_{d})$, and $I_{T}(V_{T})$ are the I-V characteristics of the
injector, detector, and thermometer tunnel junctions, respectively (see
Fig.~\ref{Fig-1}, top panel). Once the relation $T_{e}^{i}(I_{i}%
,T=\operatorname{const})$ was established, the corresponding `Thermometer'
circuit was disconnected for the rest of the experiments.

\section{Results and discussion}

\label{results}

To avoid confusion it is instructive to summarize the definitions of various
parameters of the dimensionality `temperature' hereafter used in the paper.
The $T$ stands for the bath temperature of the system measured by the two
RuO$_{x}$ resistors thermally anchored to the mixing chamber and the massive
copper sample holder. The RuO$_{x}$ sensors were calibrated against the
nuclear orientation thermometer, and during the experiments their readings
differ less than by few mK. 

The $T_{e}^{i}$ and $T_{e}^{d}$ denote the electron
temperatures of the normal metal injector and detectors, respectively. The
$T_{e}^{i}$ and $T_{e}^{d}$ are determined by fitting the corresponding
$I_{i}(V_{i})$ and $I_{d}(V_{d})$ experimental dependencies by the well-known
equilibrium expression for the tunnel current of a NIS junction. The
$T_{e}^{i}$ depends on the injection current, $T_{e}^{i}(I_{i})$, due to the
trivial Joule heating of the normal metal injector. At the lowest bath
temperature $T\simeq$20 mK and the highest injection currents $I_{i}\simeq$1
$\mu A$ the electron temperature of the injector increased by $\delta
T_{e}^{i}\equiv T_{e}^{i}-T\simeq$100 mK. The electron-phonon interaction is
rather weak at ultra-low temperatures, \cite{Pekola NIS review} and the
superconducting `body' of the sample\textbf{, }thermally anchored to the
substrate\textbf{,} has low thermal conductivity. Hence, the phonon
temperature of the remote detectors should not deviate significantly from the
bath temperature $T$. On the contrary, the electron temperature $T_{e}^{d}$ of
the normal detectors, determined by fitting the zero-injection $I_{d}%
(V_{d},I_{i}=0)$ dependency, was always higher than the base temperature $T$.
The effect originates from the poor thermal coupling typical for ultra-low
temperatures and from inevitable electron heating coming from the noisy EM
environment. At a base temperature $T\simeq$ 20 mK and zero injection
$I_{i}=0$\ the absolute value of the detector electron temperature offset
$\delta T_{e}^{d}\equiv T_{e}^{d}-T$ varied from 10 to 40 mK depending on the
coupling of the particular junction with the measuring circuit. As the
detectors are decoupled by the two NIS junctions from the remote `hot'
injector, it is reasonable to assume that for the given detector its electron
temperature $T_{e}^{d}$ should not depend on the injection current $I_{i}$:
$T_{e}^{d}(I_{i}>0,T)\simeq T_{e}^{d}(I_{i}=0,T)$. 

The $T_{e}^{S}$ denote the superconductor electron temperature. 
Formally, it should be determined taking
into account a complicated energy balance. \cite{Pekola NIS review} From the
very basic considerations one may expect that at non-zero quasiparticle
injections $I_{i}>0$ the actual (thermodynamic) temperature of the
superconductor $T_{e}^{S}$, which enters into the non-equilibrium distribution
function $f_{T}^{S}(E,T_{e}^{S})$, should be higher than the bath temperature
$T$. However, within the phenomenological formalism employed in the paper,
$T_{e}^{S}$ cannot be determined. More elaborated microscopic analysis
(to our best knowledge, currently absent) should be used to determine the
non-equilibrium (and - essentially asymmetric with respect to the chemical
potential) distribution function. Otherwise, any arbitrary value of $T_{e}%
^{S}$ substituted into a symmetric (\textit{e.g.} equilibrium) distribution function
$f_{T}^{S}(E,T_{e}^{S})$ gives exactly the same tunnel current of a NIS
junction. 

Finally, the $T^{\ast}$ characterizes the
reduction of the superconducting gap $\Delta$. When the superconducting gap is
suppressed by the quasiparticle injection one can use two complementary
descriptions: either to deal with the bare experimental data $\Delta(I_{i})$,
or to convert the non-equilibrium energy gap into an effective temperature
$T^{\ast}$ using conventional (equilibrium) BCS expression. In the latter case
$T^{\ast}$ just means: what would be the equilibrium temperature of the
superconductor $T=T^{\ast}$ resulting in the corresponding reduction of the
gap $\Delta(T_{e}^{S},I_{i}>0)=\Delta_{BCS}(T^{\ast},I_{i}=0)$. It should be
emphasized that $T^{\ast}$ is only a convenient parameter of dimensionality
`temperature' and has no direct relation to the (non-equilibrium)
thermodynamic temperature $T_{e}^{S}$. As it will be shown below, the impact
of quasiparticle injection on a superconductor is manifold. In particular, the
longitudinal (energy) mode cannot be described solely by the reduction of the
gap, or alternatively - the effective temperature $T^{\ast}$.

\begin{figure}[t]
\epsfxsize=7cm\epsffile{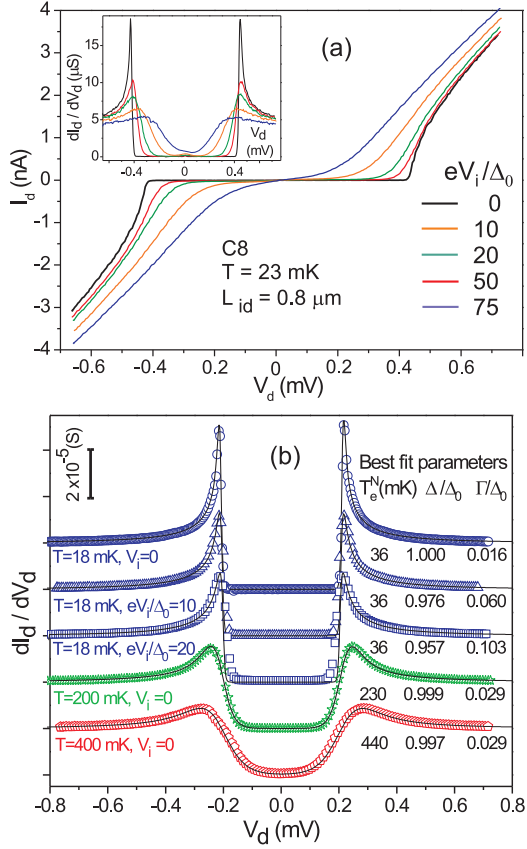}\caption{(Color online) (a) Sample C8
$I_{d}(V_{d})$ characteristics at various injections for the double junction
NISIN detector located at $L_{id}$=0.8 $\mu$m away from the injector. Inset:
the corresponding $dI_{d}/dV_{d}$ dependencies. Slight asymmetry of the
$dI_{d}/dV_{d}$ at high injections presumably originates from the not perfect
equality of the each of the NIS junctions connetced in series. (b) Sample C2
NIS detector located at $L_{id}$=2.8 $\mu$m experimental $dI_{d}/dV_{d}$
dependencies at T=18$\pm$1 mK at three injections $eV_{i}/\Delta_{i}=$ 0
($\bigcirc$), 10 ($\triangle$) and 20 ($\square$), and at $T=200$ ($\bigstar$)
and $T=400$ ($\Diamond$) mK at zero injections. Solid lines are the
theoretical best fits with the fitting parameters indicated in the plot.}%
\label{Fig-2}%
\end{figure}

Injection of non-equilibrium quasiparticles into a superconductor results in a
deviation of its density of states (DOS) $N_{S}$ and distribution functions
$f_{L,T}^{S}$ from equilibrium. Under certain assumptions the distribution
function can be found by deconvolution of the tunneling I-V dependencies.
\cite{SACLAY deconvolution} In more general cases\textbf{,} self-consistent
solution of Keldysh-Usadel equations is required. \cite{Belzig} For the
interpretation of the results we take\textbf{ }the simplified approach
\textit{postulating} equilibrium functional forms of the DOS and distribution
function $f_{S}$, but with parameters deviating from their equilibrium
values,
\begin{align}
N_{S}(E,\Delta,\Gamma)  & = \mid\operatorname{Re}\{(E+i\Gamma/2)/\sqrt{\left(
E+i\Gamma/2\right)  ^{2}-\Delta^{2}}\}\mid,\label{Eq. 1}\\
f_{S}(E,T_{e}^{S},\mu^{\ast})  & = 1/\left\{  \exp\left(  (E-\mu^{\ast}%
)/k_{B}T_{e}^{S}\right)  +1\right\} .\label{Eq. 2}%
\end{align}

Deviation of $f_{L}^{S}$ from equilibrium formally corresponds to an increase
of $T_{e}^{S}$ above the bath temperature $T$, while deviation of $f_{T}^{S}$
- to a non-zero shift of chemical potential $\mu^{\ast}$. Considering the
above expression for $f_{S}$ with two parameters, $T_{e}^{S}$ and $\mu^{\ast}%
$, one might account for both the longitudinal and the transverse mode
excitations. For the analysis of the experimental data we utilized the
simplified phenomenological approach with three fitting parameters: DOS
broadening $\Gamma$, \cite{Dynes Gamma} superconducting gap $\Delta$, and
effective quasiparticle chemical potential $\mu^{\ast}$. \cite{Parker T*,
Owen-Scalapino mu*} All parameters can be defined by fitting experimental
$I_{d}(V_{d})$ or/and $dI_{d}/dV_{d}$ dependencies. It should be noted that
the utilized phenomenological analysis, based on the postulated equilibrium
functional form of the distribution function $f_{S}(E,T_{e}^{S},\mu^{\ast}),$
disables determination of the (thermodynamic) temperature of the
superconductor $T_{e}^{S}$. More elaborated model (to our best knowledge -
currently absent) is mandatory to deduce the non-equilibrium $f_{S}%
(E,T_{e}^{S},\mu^{\ast})$ and the corresponding\ $T_{e}^{S}$.

\begin{figure}[t]
\epsfxsize=7cm\epsffile{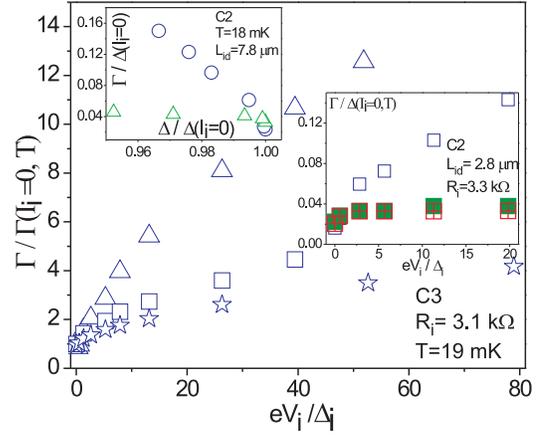}\caption{(Color online) Deparing parameter
$\Gamma$ vs. injection energy for three detectors at $L_{id}$=0.8 ($\triangle
$), 2.8 ($\square$) and 7.8 ($\bigstar$) $\mu$m. Left inset: sample C2 NIS
detector located at $L_{id}=7.8$ $\mu m$ $\Gamma(I_{i})$ vs. $\Delta(I_{i})$
at various injection currents $I_{i}$ at $T=18$ mK ($\bigcirc$) \ and $T=200$
mK ($\triangle$). Right inset: sample C2 $\Gamma$ vs. injection energy at
$L_{id}$=2.8 $\mu$m at $T=$17.5 ($\square$), 200 ($\blacksquare$) and 400 mK
($\boxplus$).}%
\label{Fig-3}%
\end{figure}

As the injected non-equilibrium quasiparticles relax on a certain length, the
fitting parameters should depend on the injection rate (energy $eV_{i}$ or
current $I_{i}=V_{i}G_{NN}^{i}$) and the distance between the detector and the
injector $L_{id}$. The $I_{d}(V_{d})$ dependency of a NIS detector in the
presence of non-equilibrium injection is
\begin{equation}
I_{d}=(G_{NN}^{d}/e)\int_{0}^{\infty}N_{S}(E,\Delta,\Gamma)\left[  f_{T}%
^{N}-f_{T}^{S}\right]  dE,\label{Eq. 3}%
\end{equation}
where $G_{NN}^{d}$ is the normal state conductivity of the junction,
$f_{T}^{N}(E,V_{d},T_{e}^{N})=\tanh((E+eV_{d})/2k_{B}T_{e}^{N})-\tanh
((E-eV_{d})/2k_{B}T_{e}^{N})$ the equilibrium value of $f_{T}$ in the normal
detector, and $f_{T}^{S}(E,L_{id},T_{e}^{S})$ the local non-equilibrium value
of $f_{T}$ in the superconductor. The first term corresponds to the text-book
equilibrium expression for a NIS tunnel current, explicitly depending on
properties of the superconductor only through its DOS $N_{S}(E,\Delta,\Gamma)$.

\begin{figure}[t]
\epsfxsize=7cm\epsffile{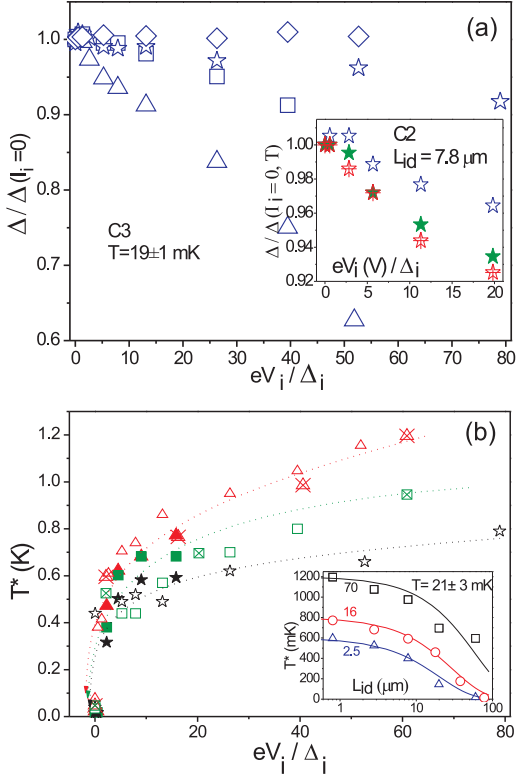}\caption{(Color online) (a) Superconducting
gap $\Delta$ vs. injection energy at $L_{id}$=0.8 ($\triangle$), 2.8
($\square$), 7.8 ($\bigstar$) and 17.8 ($\Diamond$) $\mu$m. Inset: reduction
of the superconducting gap measured by the same NIS detector at $L_{id}$=7.8
$\mu$m at $T=$17.5 (open star), 200 ($\bigstar$) and 400 (crossed star) mK.
(b) Effective temperature $T^{\ast}$ vs. injection energy for samples C3 (18
mK, open symbols), C2 (19 mK, solid symbols) and C8 (24 mK, crossed symbols)
at $L_{id}$=0.8 ($\triangle$), 2.8 ($\square$) and 7.8 ($\bigstar$) $\mu$m.
Lines are guides to the eye. Inset: spatial decay of $T^{\ast}$ averaged over
several samples at $eV_{i}/\Delta_{i}$= 70 ($\square$), 16 ($\bigcirc$) and
2.5 ($\triangle$). Lines are fits assuming exponential dependence $\sim
\exp(-L_{id}/\Lambda_{T^{\ast}})$.}%
\label{Fig-4}%
\end{figure}

The shape of the experimental $I_{d}(V_{d})$ characteristic depends on the
quasiparticle injection (Fig.~\ref{Fig-2}). By fitting $I_{d}(V_{d})$ or/and
$dI_{d}/dV_{d}$ dependencies one finds that the parameters $\Gamma$ and
$\Delta$ depend on the injection rate, proximity to the injector $L_{id}$, and
temperature $T$ (Figs.~\ref{Fig-3}, \ref{Fig-4}). The smearing of the
$I_{d}(V_{d})$ dependence [rounding of the `corners'at $eV_{d}\simeq
\Delta(I_{i})$] can be assigned to the parameter $\Gamma(I_{i})$ which enters
the superconductor DOS and is physically related to the finite quasiparticle
lifetime. \cite{Dynes Gamma} For the same injection rate $I_{i}$ the
temperature dependence $\Gamma(T, \; I_{i}= \operatorname{const}, \;
L_{id}=\operatorname{const})$ (right inset in Fig. \ref{Fig-3}) most probably
originates from the more intensive relaxation of the quasiparticles
(\textit{i.e.} from the reduction of the quasiparticle lifetime responsible
for finite $\Gamma$) due to inelastic scattering on phonons. The total number
of quasiparticles entering the superconductor per unit time is proportional to
$I_{i}$, while only the fraction $\sim\exp(-L_{id}/\sqrt{D\tau_{R}})$ reaches
the detector. At low temperatures the recombination time $\tau_R$ can be estimated
from the relation, $\tau_{0}/\tau
_{R}\sim(T/T_{c})^{1/2}\exp(-\Delta/k_{B}T)$, with $\tau_{0}$ being the
electron-phonon scattering time. \cite{Kaplan76, Chi-Clarke} These simple
arguments qualitatively explain the spatial, energy, and temperature
dependencies of $\Gamma$ (Fig.~\ref{Fig-3}), while a more elaborated model,
presumably incorporating the effect of the environment, \cite{Pekola Gamma} is
required for a quantitative analysis.

Neglecting the superconducting phase, the gap equation for the pairing
potential $\Delta$, to be solved self-consistently with the Keldysh-Usadel
equations, is $\Delta=\lambda\int_{0}^{\infty}f_{L}^{S}(E,L_{id}%
)\operatorname{Re}(F)dE$, where $\lambda$ is the electron-phonon coupling
constant, $F$ the anomalous pair amplitude, and $f_{L}^{S}(E,L_{id})$ the
local non-equilibrium value of $f_{L}$ in the superconductor. \cite{Belzig} At
any energy $E$ the function $f_{L}^{S}(E,L_{id})$ is smaller than its
equilibrium value $\tanh(E/2k_{B}T)$, \cite{VH} reducing $\Delta$
(Fig.~\ref{Fig-4}a). The effect is smaller at the lowest temperatures and
increases with increasing bath temperature (Fig.~\ref{Fig-4}a, inset).
Utilizing the well-known BCS temperature dependence for the superconducting
gap one can assign an effective temperature $T^{\ast}$, which would provide
the same effect on the energy gap as the quasiparticle injection:
$\Delta(T_{e}^{S},I_{i},L_{id})=\Delta_{BCS}(T^{\ast},I_{i}=0,L_{id})$
(Fig.~\ref{Fig-4}b). Assuming an exponential decay $T^{\ast}$=$T^{\ast}%
(0)\exp(-L_{id}/\Lambda_{T^{\ast}})$ one finds the characteristic length
$\Lambda_{T^{\ast}}=$ 40 $\mu m$ $\pm$ 20 $\mu m$ (Fig.~\ref{Fig-4}b, inset).
The large uncertainty comes from the weak $\Delta_{BCS}(T)$ dependence at low
temperatures and the limited set of $L_{id}$ points (number of detectors).
Once again we would like to remind that $T^{\ast}$ is a convenient parameter
introduced to account for the reduction of the superconducting gap
$\Delta(I_{i})$ and has no direct relation to the thermodynamic temperature of
the superconductor $T_{e}^{S}$ which should enter the corresponding
non-equilibrium distribution function. It might even happen that in the limit
of strong injection and weak quasiparticle-quasiparticle scattering $T_{e}%
^{S}$ cannot be defined at all, while the non-equilibrium energy gap
$\Delta(I_{i})$, or alternatively $T^{\ast}(I_{i})$, can be still deduced from
the experimental $I_{d}(V_{d})$ characteristics.

\begin{figure}[t]
\epsfxsize=6.5cm\epsffile{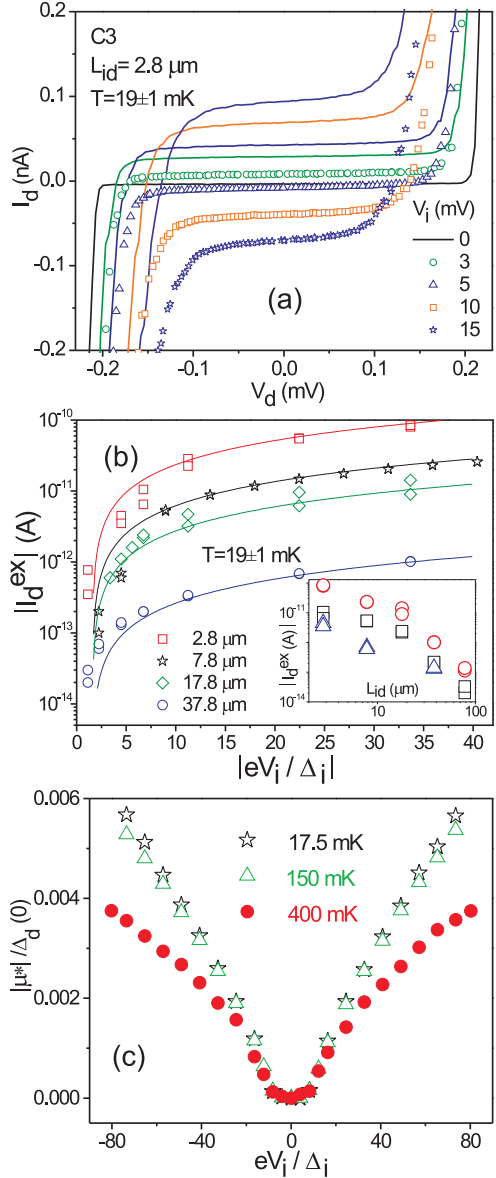}\caption{(Color online) Sample C3 NIS
detection. (a) Zoom of the sub-gap region of the $I_{d}(V_{d})$ characteristic
at different injection voltages $V_{i}$=0, 3, 5, 10 and 15 mV. Lines
correspond to the same injections of reverse polarity. (b) Excess current
$I_{d}^{ex}$ vs. injection energy $eV_{i}/\Delta_{i}$ at different distances
$L_{id}$. Lines are fits utilizing $\Lambda_{Q^{\ast}}$ = 3.6, 3.6, 5.2 and
6.4 $\mu$m for the detectors located at $L_{id}$ = 2.8, 7.8, 17.8 and 37.8
$\mu$m, \textbf{ }respectively. Inset: spatial decay of the excess current
measured at $\mid eV_{i}/\Delta_{i}\mid$=33.7 ($\bigcirc$), 6.7 ($\square$)
and 4.5($\triangle$). (c) Normalized chemical potential $\mu^{\ast}$ vs.
injection energy measured at various temperatures by the same detector at
$L_{id}$=7.8 $\mu$m.}%
\label{Fig-5}%
\end{figure}

We believe that the reduction of the gap $\Delta(I_{i})$ and the increase of
$\Gamma(I_{i})$\ are the two manifestations of the same phenomenon - energy
disbalance (longitudinal mode) accompanying injection of the non-equilibrium
quasiparticles.\ Presumably, in a comprehensive microscopic description (to
our best knowledge, currently absent) the longitudinal and the transverse
modes may be entangled both contributing to the $\Delta(I_{i})$ and
$\Gamma(I_{i})$ dependencies. Within the `orthodox' BCS model the
(temperature-dependent) smearing of a NIS detector $I_{d}(V_{d})$
characteristic comes solely from the distribution function of the normal
electrode $f^{N}(E,V_{d},T_{e}^{N})$ and not from the DOS or distribution
function of the superconductor. Temperature of the superconductor $T_{e}^{S}$
enters solely through the temperature dependence of the gap $\Delta(T_{e}%
^{S})$. In a typical equilibrium situation $T=T_{e}^{N}=T_{e}^{S}$. Employed
in the paper analysis deviates from that `orthodox' model by introduction of
the Dynes smearing $\Gamma\lbrack I_{i}]$ entering the DOS of the
superconductor $N_{S}(E,\Delta\lbrack I_{i}],\Gamma\lbrack I_{i}])$,
Eq.~\eqref{Eq. 1}, at a given, but not defined, temperature of the
superconductor $T_{e}^{S}$, which indeed leads to the extra (non-temperature
dependent) smearing of the $I_{d}(V_{d})$ dependence, Eq.~\eqref{Eq. 3}.

The employed analysis of the experimental data based on the equilibrium
expression for the tunnel current [first term in Eq.~\eqref{Eq. 3}] with the
DOS deviating from the `orthodox' BCS model [Eqs.~\eqref{Eq. 1},
\eqref{Eq. 2}] enables separation of the two impacts $\Delta(I_{i})$ and
$\Gamma(I_{i})$. The reduction of the gap $\Delta(I_{i})$ (Fig. \ref{Fig-4})
leads to a `sharp' $I_{d}(V_{d})$ characteristic: as if the temperature is
still low $T_{e}^{S}\simeq T$, but the energy gap is reduced $\Delta
(I_{i}>0)<\Delta(I_{i}=0)$. The impact of the finite $\Gamma(I_{i})$ manifests
itself as the smearing of the $I_{d}(V_{d})$ characteristics at the `corners'
$eV_{d}\simeq\pm\Delta(I_{i})$ which cannot be accounted for any reasonable
value of $T_{e}^{N}$.

It is important to emphasize that the experimental non-equilibrium
$I_{d}(V_{d},I_{i}>0,T)$ dependencies cannot be described by the `orthodox'
BCS model just assuming that the temperature of the system has increased above
the bath temperature: $T\rightarrow T_{e}^{N},T_{e}^{S}\gg T$. At $T\ll T_{c}%
$\ the shape of the experimental $I_{d}(V_{d},I_{i}>0,T)$ characteristics is
qualitatively different from the reference equilibrium characteristics
$I_{d}(V_{d},I_{i}=0,T^{\prime})$ taken at elevated temperatures $T^{\prime
}>T$ (Fig. \ref{Eq. 2}b). Those zero-injection high-temperature $I_{d}(V_{d})$
characteristics can be nicely fit by the equilibrium BCS expression for the
tunnel current [first term in Eq.~\eqref{Eq. 3}] assuming small intrinsic
Dynes broadening of the DOS, Eq.~\eqref{Eq. 1}, with $\Gamma(I_{i}%
=0)/\Delta(I_{i}=0)\simeq$ 0.02 (Fig. \ref{Eq. 2}b) in a reasonable agreement
with existing data on aluminum. \cite{Pekola Gamma} However, one runs into a
contradiction trying to fit the experimental $I_{d}(V_{d},I_{i}>0,T)$
characteristics using a single fitting parameter $T_{fit}$ and the first term
in the expression \eqref{Eq. 3} for the NIS tunnel current. To properly
account for the gap reduction $\Delta(I_{i})$ one should set the best fit
`equilibrium' temperature $T_{fit}=T^{\ast}\gg T$ (compare Figs. \ref{Fig-2}b
and \ref{Fig-4}b) being significantly higher than the temperature of `hottest'
part of the system - the injector. While to account for the observed
broadening at the gap edge, the effective fitting temperature $T_{fit}$ should
be set significantly lower $T_{fit}\simeq T_{N}^{e}\ll T^{\ast}$ (Fig.
\ref{Fig-2}b). Summarizing, one should conclude that at non-zero injections
the $I_{d}(V_{d},I_{i}>0,T)$ characteristics cannot be described by a single
`equilibrium' temperature $T_{fit}$. The two independent parameters,
$\Delta(I_{i})$ and $\Gamma(I_{i})$,\ are necessary. The dependencies
$\Delta(I_{i})$ and $\Gamma(I_{i})$ on the injection rate are different. At
ultra-low temperatures $T\ll T_{c\text{ }}$and modest injections
$eV_{i}/\Delta_{0}\lesssim$10 the suppression of the gap is rather small (Fig.
\ref{Fig-4}a), while the increase of the DOS broadening parameter $\Gamma$\ is
quite pronounced (Fig. \ref{Fig-3}). At elevated temperatures the dependence
of $\Gamma$\ on the injection rate is rather weak (insets in Fig.
\ref{Fig-3}). We would like to emphasize that the necessity to introduce the
two independent parameters $\Delta(I_{i})$ and $\Gamma(I_{i})$ is the result
of the employed phenomenological approach based on
Eqs.~\eqref{Eq. 1}-\eqref{Eq. 3}. We would wish to believe that in an
elaborated microscopic model dealing with the essentially non-equilibrium
distribution function $f_{S}(E,T_{e}^{S})$ a single parameter - the truly
thermodynamic temperature of the superconductor $T_{e}^{S}$ - would be
sufficient to account for the whole variety of the experimental data.

A remarkable feature of the tunnel current expression is the existence of the
non-zero excess current $I_{d}^{ex}=I_{d}(V_{d}=0)$ originating from the
independence of $f_{T}^{S}(E,L_{id},T_{e}^{S})$ on the bias volatage $V_{d}$
[second term in Eq.~\eqref{Eq. 3}]. The effect has been observed in NISIS'
flat sandwiches \cite{Clarke PRL72, Tinkham-Clarke I-V PRL72} and recently in
multiterminal NIS structures. \cite{Yagi} Within our phenomenological
approach, the experimentally measured excess current is linked to the
effective quasiparticle chemical potential, \cite{Tinkham-Clarke I-V PRL72}
$I_{d}^{ex}=\left(  G_{NN}^{d}/e\right)  \mu^{\ast}$. A typical example of a
NIS detector sub-gap $I_{d}(V_{d})$ characteristic is shown in
Fig.~\ref{Fig-5}a. The excess current increases with increase of the injection
rate reversing its sign with the change of the injection polarity. The effect
decreases with increasing distance to the injector $L_{id}$ (Fig.~\ref{Fig-5}%
b) and the bath temperature (Fig.~\ref{Fig-5}c). Slight asymmetry of
$I_{d}^{ex}$ with respect to zero injection (Fig.~\ref{Fig-5}a) probably
originates from a parasitic residual thermo-electric potential. It should be
noted that in the NISIN detector geometry the excess current is absent since
the contributions from the two NIS junctions connected in series and located
at the same distance $L_{id}$ (from the opposite sides of the superconducting
bar, see Fig.~\ref{Fig-1}), cancel each other. The double NISIN junctions were
used only as a complementary configuration in the experiments on the energy
mode relaxation (Fig.~\ref{Fig-2}a).

The solution of the diffusion equation for the charge imbalance $Q^{\ast}$
gives the excess current
\begin{equation}
I_{d}^{ex}=I_{i}\frac{F^{\ast}\Lambda_{Q^{\ast}}G_{NN}^{d}}{2e^{2}N(0)D\sigma
}\exp(-L_{id}/\Lambda_{Q^{\ast}}),
\end{equation}
where $N(0)=1.08\times10^{47}$ 1/J$\times$m$^{3}$ is the normal state DOS for
aluminium at the Fermi level, and $F^{\ast}$ is a smooth function varying from
0 at no injection to 1 at $eV_{i}$ $>>\Delta_{i}$. \cite{Tinkham-Clarke I-V
PRL72} Substituting the mean free path $\ell\simeq20$ nm, Fermi velocity
$v_{F}=1.36\times10^{6}$ m/s, diffusivity $D=1/3v_{F}\ell$ and cross section
$\sigma=25$ nm $\times$ $400$ nm one gets an acceptable fit for the dependence
of the excess current on the injection energy (Fig.~\ref{Fig-5}b). The
best-fit charge imbalance length $\Lambda_{Q^{\ast}}=\sqrt{D\tau_{Q^{\ast}}}$
varies from 3.5 to 6.5 $\mu$m being in a reasonable agreement with recent
findings. \cite{Yagi, Beckmann Lambda*}

Weak temperature dependence of the charge imbalance relaxation $Q^{\ast}\sim
I_{d}^{ex}\sim\mu^{\ast}$ (Fig.~\ref{Fig-5}c) at ultra-low temperatures is
expected: in the absence of an effective electron-phonon interaction the
remaining (weak) mechanism is the elastic scattering due to the gap
inhomogeneity, or/and anisotropy or even just a finite supercurrent.
\cite{Tinkham-Clarke I-V PRL72} Contrary to the transverse mode relaxation,
the relaxation of the longitudinal (energy) mode always requires the presence
of inelastic mechanism(s). At the ultra-low temperatures of the experiment
$k_{B}T\ll\Delta$ there are very few equilibrium phonons capable to provide
the quasiparticle energy relaxation. The only (potential) source of the
phonons with energies above the bath temperature $T$ is the Joule heating in
the injector. However, how it has been already discussed, even in the
worst-case scenario of the strongest injection the diffusion of heat from the
`hot' injector into the superconductor is not sufficient to account for our
findings. The maximum rise of the electron temperature in the injector $\delta
T_{e}^{i} \simeq$ 100 mK. Even assuming a perfect electron-phonon coupling in
the injector, the energy of the corresponding phonons is still much smaller
than the energy $\geq2\Delta$ required for a pair of hole-like and
electron-like quasiparticles to form a Cooper pair. However, as the
quasiparticle recombination finally does happen on `astronomically' large (for
a superconductor) scales $\Lambda_{T^{\ast}}\simeq$ 40 $\mu$m, some weak
mechanism should exist. For example, emission / re-absorption of
non-equilibrium phonons or photons, might provide the required energy
relaxation. Certainly, these mechanisms should also contribute to the
relaxation of the transverse mode. The observation $\Lambda_{Q^{\ast}}%
<\Lambda_{T^{\ast}}$ supports our conclusion that at ultra-low temperatures
the elastic scattering is an important channel of the charge imbalance relaxation.

Substantial quantitative difference between $\Lambda_{T^{\ast}}$ and
$\Lambda_{Q^{\ast}}$ at ultra-low temperatures, $T\ll T_{c}$, is in a
qualitative agreement with previous calculations. \cite{Kaplan76, Chi-Clarke}
The explicit condition when one can neglect the elastic scattering has been
obtained in Ref.~\onlinecite{Bezugli-Bratus-Galaiko-FNT77}, $(T_{c}%
-T)/T_{c}\ll(\tau_{E}T_{c})^{-2/3}$, where $\tau_{E}$ is the energy relaxation
time (measured in $1/K$). Analysis of this expression leads to the conclusion
that the only physical limit when always $\Lambda_{Q^{\ast}}\gg$ $\Lambda_{E}$
is at $T\rightarrow T_{c}$, where $\Lambda_{Q^{\ast}}$ diverges. Formally it
means that in the normal state there is no branch relaxation mechanism. Our
experiments were made at very low temperatures much smaller then $T_{c}$. At
such low temperatures, to our best knowledge, there is no general criterion to
\textit{a priori} relate $\Lambda_{Q^{\ast}}$ with $\Lambda_{T^{\ast}}$.

\section{Conclusion}

\label{secconcl}

In conclusion, we performed spatially resolved measurements of the
non-equilibrium quasiparticle relaxation in superconducting aluminum at
ultra-low temperatures $T\ll T_{c}$. The analysis employs the phenomenological
interpretation postulating equilibrium functional dependencies of DOS and
distribution function of the superconductor, while assuming that the
broadening parameter $\Gamma$, effective chemical potential $\mu^{\ast}$, and
superconducting gap $\Delta$ depend on the rate of quasiparticle injection and
the proximity to the injector. Two scales $\Lambda_{T^{\ast}}=$ 
$40\pm20$ $\mu$m and $\Lambda_{Q^{\ast}}$ $=5\pm1.5$ $\mu$m are responsible for
relaxation of the energy (longitudinal) and charge (transverse) modes,
respectively. Both quantities were measured simultaneously in the same samples
using the same experimental technique eliminating sample / measurement
artifacts. This is the central result of the paper. It should be emphasized
that both the energy and the charge disequilibrium are universal phenomena
which should be taken into consideration in a broad class of systems involving
electron, \cite{Pekola NIS review, Arutyunov_cooling, Vasenko_cooling, 
Kuzmin bolometer,SET poisoning} spin and/or
coherent non-local \cite{non-local and spintronic} transport. Despite the
reasonable agreement with the phenomenological description, application of the
relaxation approximation for the essentially spatially inhomogeneous problem
is not fully justified at low temperatures $T\ll T_{c}$. A more elaborated
microscopic model is required for a quantitative analysis. We hope that our
findings will trigger the corresponding research activity.

\begin{acknowledgments}
The authors would like to acknowledge support of TEKES project DEMAPP, FCPK
grant 2010-1.5-508-005-037, NanoSCIERA project Nanofridge and ANR-DYCOSMA. We
thank A. Julukian and L. Leino for AFM analysis, and F. Hekking for
discussions. A.S.V. acknowledge the hospitality of Donostia International
Physics Center (DIPS) during his stay in San Sebasti\'{a}n, Spain.
\end{acknowledgments}


\end{document}